\pdfoutput=1
\RequirePackage{ifpdf}
\ifpdf 
\documentclass[pdftex]{sigma}
\else
\documentclass{sigma}
\fi

\newtheorem{Lemma}{Lemma}
{\theoremstyle{definition}\newtheorem{Notation}{Notation}
\newtheorem{Remark}{Remark}
}

\DeclareMathOperator{\num}{num}
\DeclareMathOperator{\den}{den}

\numberwithin{Notation}{section}
\numberwithin{Remark}{section}
\numberwithin{Lemma}{section}
\numberwithin{equation}{section}
\begin{document}

\allowdisplaybreaks

\renewcommand{\PaperNumber}{059}

\FirstPageHeading

\ShortArticleName{Solvable Many-Body Models of Goldf\/ish Type with One-, Two- and Three-Body Forces}

\ArticleName{Solvable Many-Body Models of Goldf\/ish Type\\
with One-, Two- and Three-Body Forces}

\Author{Oksana BIHUN~$^\dag$ and Francesco CALOGERO~$^\ddag$}

\AuthorNameForHeading{O.~Bihun and F.~Calogero}

\Address{$^\dag$~Department of Mathematics, Concordia College at Moorhead, MN, USA}
\EmailD{\href{mailto:obihun@cord.edu}{obihun@cord.edu}}
\URLaddressD{\url{http://wwwp.cord.edu/faculty/obihun/Oksana\%20Bihun.htm}}

\Address{$^\ddag$~Physics Department, University of Rome ``La Sapienza'', Istituto Nazionale di Fisica Nucleare,\\
\hphantom{$^\ddag$}~Sezione di Roma, Italy}
\EmailD{\href{mailto:francesco.calogero@roma1.infn.it}{francesco.calogero@roma1.infn.it},
\href{mailto:francesco.calogero@uniroma1.it}{francesco.calogero@uniroma1.it}}

\ArticleDates{Received June 07, 2013, in f\/inal form October 02, 2013; Published online October 09, 2013}

\Abstract{The class of \textit{solvable} many-body problems ``of goldf\/ish type'' is extended by including
(the additional presence of) \textit{three-body} forces.
The solvable~$N$-body problems thereby identif\/ied are characterized by Newtonian equations of motion
featuring 19 arbitra\-ry ``coupling constants''.
Restrictions on these constants are identif\/ied which cause these systems~-- or appropriate variants of
them~-- to be \textit{isochronous} or \textit{asymptotically isochronous}, i.e.\ \textit{all} their
solutions to be \textit{periodic} with a~\textit{fixed} period (independent of the initial data) or to have
this property up to contributions vanishing exponentially as $t\rightarrow \infty $.}


\Classification{70F10; 70H06; 37J35; 37K10}

\section{Introduction}

Over three decades ago~\cite{C1978} a~class of solvable~$N$-body problems featuring several free parameters
(``coupling constants'') was introduced by identifying the coordinates $z_{n}(t)$ of the moving particles
with the~$N$ zeros of the time-dependent (monic) polynomial $\psi(z,t)$ (of degree~$N$ in the independent
variable~$z$),
\begin{gather}
\psi(z,t)=\prod\limits_{n=1}^{N}[z-z_{n}(t)]=z^{N}+\sum_{m=1}^{N}c_{m}(t)z^{N-m},
\label{psizt}
\end{gather}
itself evolving according to a~\textit{linear} partial dif\/ferential equation (PDE)~-- suitably restricted
to guarantee that it feature a~polynomial solution of degree~$N$ in~$z$.
The simplest dynamical system belonging to this class displays remarkably neat properties and was therefore
considered a~``goldf\/ish'' (for a~justif\/ication of this terminology, see~\cite{Gold}; subsequently this
terminology has been often employed to identify~$N$-body models belonging to this class, and this
justif\/ies its use also in the present paper, including its title).
The behavior of the solutions of these~$N$-body problems has been variously investigated and also used to
arrive at related mathematical results, such as f\/inite-dimensional representations of dif\/ferential
operators and Diophantine properties of the zeros of certain polynomials: see the two
monographs~\cite{C2001} and~\cite{C2008} and the references quoted there (including the more recent ones
added to the 2012 paperback version of~\cite{C2008}), and the more recent
papers 
\cite{BCCircle13,BCY13,C11,C12_1,C12_2,C12_3,C13_2,C13_3,C2013,CY12_3, CY12_1,CY12_2,CY12_4}.

The class of \textit{linear} PDEs satisf\/ied by the polynomial $\psi(z,t)$ which subtended these
investigations was so far restricted to PDEs featuring derivatives up to \textit{second} order.
The restriction to \textit{time}-dif\/ferentiations of \textit{second order} was motivated by the interest
in~$N$-body problems featu\-ring equations of motion of \textit{Newtonian} type (i.e., ``acceleration
equals force''); the more general restriction to dif\/ferentiations of at most \textit{second-order}
implied that the~$N$-body problems under consideration only involved \textit{one-body} and
\textit{two-body} (generally velocity-dependent) forces.
In the present paper we extend our consideration to \textit{third-order}~$z$-derivatives~-- while
maintaining the restriction to \textit{second-order time}-dif\/ferentiations so as to only treat~$N$-body
problems of \textit{Newtonian} type.
This entails that the corresponding \textit{solvable}~$N$-body problems thereby identif\/ied involve,
additionally, \textit{three-body} forces.

In the following Section~\ref{section2} we report our main results, namely we display the Newtonian
equations of motion of the~$N$-body problems identif\/ied in this paper and we indicate how they are
\textit{solved by algebraic operations}.
In a~terse Section~\ref{section3} these f\/indings are proven: these developments rely on identities
reported and proven in Appendix~\ref{appendixA}, constituting a~substantial part of this paper.
In Section~\ref{section4} we highlight some special cases in which these~$N$-body problems~-- or
appropriate variants of them~-- are \textit{isochronous} or \textit{asymptotically isochronous},
i.e.~\textit{all} their solutions are \textit{periodic} with a~\textit{fixed} period (independent of the
initial data) or they feature this property up to contributions vanishing exponentially as $t\rightarrow
\infty $.
A terse Section~\ref{section5} entitled ``Outlook'' outlines possible future developments.

\section{Main results}
\label{section2}

In this section we report the main results of this paper, which are then proven in the following section.
But f\/irstly let us specify our notation.
\begin{Notation}\looseness=-1
\label{Notation2.1}
The~$N$ coordinates (``dependent variables'') of the point, unit-mass, moving particles which are the
protagonists of the~$N$-body problem treated in this paper are generally denoted as $z_{n}\equiv z_{n}(t)$,
with~$t$ (``time'') the ``independent variable''.
As usual dif\/ferentiation with respect to time is denoted by a~superimposed dot, hence $\dot{z}_{n}\equiv
dz_{n}/dt$, $\ddot{z}_{n}\equiv d^{2}z_{n}/dt^{2}$.
As we just did here, often the indication of the~$t$-dependence is not explicitly displayed.
We generally assume that these coordinates $z_{n}$ are \textit{complex} numbers, so that the points with
coordinates $z_{n}(t)$ move in the \textit{complex}~$z$-plane; but special subcases in which the
coordinates $z_{n}$ are \textit{real} are of course possible; and it is of course also possible to
reinterpret motions taking place in the \textit{complex}~$z$-plane as instead taking place in the
\textit{real} horizontal plane by identifying the real and imaginary parts of the \textit{complex} numbers
$z_{n}=x_{n}+iy_{n}$ as the Cartesian components of the \textit{real} $2$-vectors
$\vec{r}_{n}=(x_{n},y_{n})$ (as explained in Chapter~4 of~\cite{C2001}, entitled ``Solvable and/or
integrable many-body problems in the plane, obtained by complexif\/ication'').
Here and hereafter~$i$ is the imaginary unit, $i^{2}=-1$,~$N$ is an \textit{arbitrary} positive integer
(generally $N\geq 2$), and it is understood that subscripts such as $n$ (and also $m$, $k$, $\ell$, but
not~$j$; see below) run over the positive integers from~$1$ to~$N$ (unless otherwise specif\/ied); the
reader is often (but not always) explicitly reminded of this fact.
We also use occasionally the notation~$\underline{z}$ to denote the (generally \textit{complex})~$N$-vector
of components~$z_{n}$, $\underline{z}\equiv(z_{1},\dots,z_{N})$; and likewise for other underlined letters
(see below).
A key role in our treatment is played by the time-dependent (monic) polynomial~$\psi(z,t)$ of degree~$N$
in the (\textit{scalar}, generally \textit{complex}) variable~$z$ which features the~$N$ coordinates~$z_{n}(t)$ as its~$N$ zeros, see~\eqref{psizt}.
\end{Notation}

The \textit{solvable}~$N$-body problem treated in this paper is characterized by the following Newtonian
equations of motion:
\begin{gather}
\ddot{z}_{n}+E\dot{z}_{n}
=B_{0}+B_{1}z_{n}-(N-1)[2A_{3}+3(N-2)F_{4}]z_{n}^{2}+(N-1)(N-2)G_{3}\dot{z}_{n}z_{n}\notag
\\
\phantom{\ddot{z}_{n}+E\dot{z}_{n}=}
+\sum_{\ell=1;\,\ell\neq n}^{N}\big\{(z_{n}-z_{\ell})^{-1}\big[2\dot{z}
_{n}\dot{z}_{\ell}+2\big(A_{0}+A_{1}z_{n}+A_{2}z_{n}^{2}+A_{3}z_{n}^{3}\big)\notag
\\
\phantom{\ddot{z}_{n}+E\dot{z}_{n}=}
-(\dot{z}_{n}+\dot{z}_{\ell})(D_{0}+D_{1}z_{n})-D_{2}z_{n}(\dot{z}_{n}
z_{\ell}+\dot{z}_{\ell}z_{n})\big]\big\}\notag
\\
\phantom{\ddot{z}_{n}+E\dot{z}_{n}=}
+\sum_{k,\ell=1;\,k\neq n,\, \ell\neq n,\, k\neq\ell}^{N}\bigg[\frac{3(F_{0}
+F_{1}z_{n}+F_{2}z_{n}^{2}+F_{3}z_{n}^{3}+F_{4}z_{n}^{4})}{(z_{n}-z_{\ell})(z_{n}-z_{k})}\label{EqMot}
\\
\phantom{\ddot{z}_{n}+E\dot{z}_{n}=}
-\left(\frac{G_{0}+G_{1}z_{n}+G_{2}z_{n}^{2}+G_{3}z_{n}^{3}}{z_{n}
-z_{\ell}}\right)\left(\frac{\dot{z}_{n}+\dot{z}_{k}}{z_{n}-z_{k}}+\frac{\dot{z}_{\ell}+\dot{z}_{k}}
{z_{\ell}-z_{k}}\right)\bigg], \qquad
n=1,\dots,N.\notag
\end{gather}
Here and hereafter the 19 upper-case letters $A_{j}$, $B_{j}$, $D_{j}$, $E$, $F_{j}$, $G_{j}$ denote
time-independent parameters (``coupling constants'').
They are \textit{a priori arbitrary} (possibly \textit{complex}) numbers; special assignments are
occasionally considered below.
This notation is chosen consistently with equation~(2.3.3-2) of~\cite{C2001} (with $C=1$), to which this system
of~$N$~equations of motion reduces when the 9 ``new'' parameters vanish, i.e.
\begin{gather}
F_{0}=F_{1}=F_{2}=F_{3}=F_{4}=G_{0}=G_{1}=G_{2}=G_{3}=0.
\label{FGzero}
\end{gather}
Note that the terms associated with the $4$~new parameters $F_{0}$, $F_{1}$, $F_{2}$, $F_{3}$, represent
\textit{velocity-independent three-body} forces, the terms associated with the $3$~new parameters~$G_{0}$,
$G_{1}$, $G_{2}$ represent \textit{velocity-dependent three-body} forces, and the terms associated with the
new parameters~$F_{4}$~respectively $G_{3}$~represent \textit{velocity-independent} respectively
\textit{velocity-dependent} \textit{one-body} and \textit{three-body} forces.
(Of course terms representing \textit{one-body} forces can always be absorbed in those representing
\textit{many-body} forces: for instance the \textit{one-body} term $-2(N-1)A_{3}z_{n}^{2}$~in the
right-hand side of~\eqref{EqMot} could be eliminated by replacing $z_{n}^{3}$ with $z_{n}^{2}z_{\ell }$~in
the \textit{two-body} term multiplying $A_{3}$~in the f\/irst sum in the right-hand side
of~\eqref{EqMot}).

This Newtonian~$N$-body system is \textit{solvable} by \textit{algebraic} operations because~-- as shown in
the following Section~\ref{section3}~-- the coordinates $ z_{n}(t)$ evolving according to this
system of~$N$ Ordinary Dif\/ferential Equations (ODEs) coincide with the~$N$ zeros of the time-dependent
polynomial~\eqref{psizt} of degree~$N$ in~$z$, itself evolving according to the following \textit{linear}
PDE:
\begin{gather}
\psi_{tt}+\{E-(N-1)[D_{2}+(N-2)G_{3}]z\}\psi_{t}\notag
\\
\hphantom{\psi_{tt}}{}
+\big(G_{0}+G_{1}z+G_{2}z^{2}+G_{3}z^{3}\big)\psi_{zzt}+\big(D_{0}+D_{1}z+D_{2}z^{2}\big)\psi_{zt}
\notag
\\
\hphantom{\psi_{tt}}{}
+\big(F_{0}+F_{1}z+F_{2}z^{2}+F_{3}z^{3}+F_{4}z^{4}\big)\psi_{zzz}+\big(A_{0}+A_{1}z+A_{2}z^{2}+A_{3}
z^{3}\big)\psi_{zz}\notag
\\
\hphantom{\psi_{tt}}{}
+\big\{B_{0}+B_{1}z-(N-1)[2A_{3}+3(N-2)F_{4}]z^{2}\big\}\psi_{z}\notag
\\
\hphantom{\psi_{tt}}{}
+ N\big\{-B_{1}-(N-1)[A_{2}+(N-2)F_{3}]+(N-1)[A_{3}+2(N-2)F_{4}]z\big\}\psi=0.
\label{PDEpsi}
\end{gather}
Here the 19 upper-case letters are of course the same time-independent parameters featured by the Newtonian
equations of motion~\eqref{EqMot}.
Again, this notation is consistent with that used in~\cite{C2001}: indeed this PDE is a~natural
generalization of equation~(2.3.3-1) of~\cite{C2001} (with $C=1$) to which it clearly reduces when the
\textit{new} parameters vanish, see~\eqref{FGzero}.

This PDE implies that the~$N$ coef\/f\/icients $c_{m}\equiv c_{m}(t)$, see~\eqref{psizt}, evolve according
to the following system of~$N$ \textit{linear} ODEs:
\begin{gather}
\ddot{c}_{m}+(N+2-m)(N+1-m)G_{0}\dot{c}_{m-2}+(N+1-m)[D_{0}+(N-m)G_{1}]\dot{c}_{m-1}\notag
\\
\hphantom{\ddot{c}_{m}}{}
+\{E+(N-m)[D_{1}+(N-1-m)G_{2}]\}\dot{c}_{m}-m[D_{2}+(2N-3-m)G_{3}]\dot{c}_{m+1}\notag
\\
\hphantom{\ddot{c}_{m}}{}
+(N+3-m)(N+2-m)(N+1-m)F_{0}c_{m-3}\notag
\\
\hphantom{\ddot{c}_{m}}{}
+(N+2-m)(N+1-m)[A_{0}+(N-m)F_{1}]c_{m-2}\notag
\\
\hphantom{\ddot{c}_{m}}{}
+(N+1-m)\big\{(N-m)[A_{1}+(N-1-m)F_{2}]+B_{0}\big\}c_{m-1}\notag
\\
\hphantom{\ddot{c}_{m}}{}
-m\big\{(2N-1-m)A_{2}+B_{1}+\big[3N^{2}-6N+2-3(N-1)m+m^{2}\big]F_{3}\big\}c_{m}\notag
\\
\hphantom{\ddot{c}_{m}}{}
+m(m+1)[A_{3}+(3N-5-m)F_{4}]c_{m+1}=0,
\qquad
m=1,\dots,N,
\label{ODEscm}
\end{gather}
of course with $c_{0}=1$ and $c_{m}=0$ for $m<0$ and for $m>N$.

\begin{Remark}\label{Remark2.1}
This system of~$N$ ODEs satisf\/ied by the~$N$ coef\/f\/icients $c_{m}(t)$ justif\/ies the assertion made
above that the linear PDE~\eqref{PDEpsi} admits a~polynomial solution: specif\/ically, the
polynomial~\eqref{psizt} of degree~$N$ in~$z$.
Of course this system of~$N$ ODEs reduces to equation~(2.3.3-8) of~\cite{C2001} (with $C=1$) when the
\textit{new} parameters vanish, see~\eqref{FGzero}.
Likewise, it reduces to equation~(4.54) of~\cite{C2008} (up to the correction of a~trivial misprint in that
equation, and to an obvious notational change).
Note however that the analogous equation has been wrongly reported (as equation~(3)) in~\cite{C2013}: due to
a~trivial misprint (a multiplicative factor $c_{m-2}$ missing in the f\/irst term in the second line) and
the mistake of inserting two terms in the right-hand side (which should instead just be zero); fortunately
this mistake has no consequence on the remaining part of that paper~-- except for the mistaken Remark~1.1
which should of course be ignored.
\end{Remark}

This system,~\eqref{ODEscm}, of~$N$ \textit{autonomous linear} ODEs is of course solvable by
\textit{algebraic} operations.
Indeed its \textit{general} solution reads
\begin{gather}
\underline{c}(t)=
\sum_{m=1}^{N}\Bigl[b_{m}^{(+)}\underline{v}^{(+)(m)}\exp\big(\lambda_{m}^{(+)}t\big)+b_{m}
^{(-)}\underline{v}^{(-)(m)}\exp\big(\lambda_{m}^{(-)}t\big)\Bigr],
\label{Solc}
\end{gather}
where   the~$N$-vector $\underline{c}(t)$ has the~$N$ components $ c_{m}(t)$, the $2N$ (time-independent)
coef\/f\/i\-cients $ b_{m}^{(\pm)}$ are \textit{a priori} arbitrary~-- to be f\/ixed \textit{a
posteriori} in order to satisfy the $2N$ initial conditions~$ c_{m}( 0)$ and~$\dot{c}_{m}( 0)$~-- while
the $2N$ (time-independent)~$N$-vectors $\underline{v}^{(\pm) ( m)}$ respectively the $2N$
(time-independent) numbers~$\lambda_{m}^{(\pm)}$ are the $2N$ eigenvectors, respectively the $2N$
eigenvalues, of the following (time-independent) generalized matrix-vector eigenvalue problem:
\begin{subequations}
\begin{gather}
\big(\lambda^{2}\underline{I}+\lambda\underline{U}+\underline{V}\big)\underline{v}=0.
\label{EigenEq}
\end{gather}
Here of course $\underline{I}$ is the $N\times N$ unit matrix ($ I_{mn}=\delta_{m,n}$ where, here and
below, $\delta_{m,n}$ is the Kronecker symbol) and the two $N\times N$ matrices $\underline{U}$ and $
\underline{V}$ are def\/ined componentwise as follows:
\begin{gather}
U_{mn}=(N+1-m)(N+2-m)G_{0}\delta_{n,m-2}+(N+1-m)[D_{0}+(N-m)G_{1}]\delta_{n,m-1}\notag
\\
\phantom{U_{mn}=}{}
+\{E+(N-m)[D_{1}+(N-1-m)G_{2}]\}\delta_{n,m}\notag
\\
\phantom{U_{mn}=}{}
-m[D_{2}+(2N-3-m)G_{3}]\delta_{n,m+1},
\qquad
n,m=1,\dots,N;
\\
V_{mn}=(N+3-m)(N+2-m)(N+1-m)F_{0}\delta_{n,m-3}\notag
\\
\phantom{V_{mn}=}{}
+(N+2-m)(N+1-m)[A_{0}+(N-m)F_{1}]\delta_{n,m-2}\notag
\\
\phantom{V_{mn}=}{}
+(N+1-m)\big\{(N-m)[A_{1}+(N-1-m)F_{2}]+B_{0}\big\}\delta_{n,m-1}\notag
\\
\phantom{V_{mn}=}{}
-m\big\{(2N-1-m)A_{2}+B_{1}+\big[3N^{2}-6N+2-3(N-1)m+m^{2}\big]F_{3}\big\}\delta_{n,m}\notag
\\
\phantom{V_{mn}=}{}
+m(m+1)[A_{3}+(3N-5-m)F_{4}]\delta_{n,m+1}
\qquad
n,m=1,\dots,N.
\end{gather}

This implies of course that the $2N$ eigenvalues $\lambda_{m}^{(\pm)}$ are the $2N$ roots of the
following polynomial equation (of degree $2N$ in $\lambda $):
\begin{gather}
\det\big(\lambda^{2}\underline{I}+\lambda\underline{U}+\underline{V}\big)=0.
\label{deter}
\end{gather}
\end{subequations}

These f\/indings show that the solution of the system~\eqref{ODEscm} is achieved by the \textit{algebraic}
operation of determining the eigenvalues and eigenvectors of the matrix-vector generalized eigenvalue
problem~\eqref{EigenEq}.

The algebraic equation~\eqref{deter} can be \textit{explicitly} solved (for arbitrary~$N$) in the two
special cases in which the two $N\times N$ matrices $\underline{U}$ and $\underline{V}$ are either both
\textit{upper triangular} or both \textit{lower triangular}.

The \textit{first} of these two special cases obtains if, of the 19 parameters in~\eqref{ODEscm}, the
following 4 vanish:
\begin{gather}
A_{3}=D_{2}=F_{4}=G_{3}=0.
\label{Res1}
\end{gather}

The \textit{second} of these two special cases obtains if instead, of the 19 parameters in~\eqref{ODEscm},
the following 9 vanish:
\begin{gather}
A_{0}=A_{1}=B_{0}=D_{0}=F_{0}=F_{1}=F_{2}=G_{0}=G_{1}=0.
\label{Res2}
\end{gather}

It is then easily seen that~-- in both these two cases~-- the $2N$ eigenvalues $\lambda_{m}^{(\pm)}$ read
\begin{subequations}
\label{landas}
\begin{gather}
\lambda_{m}^{(\pm)}=-\frac{1}{2}\big\{E+(N-m)[D_{1}+(N-1-m)G_{2}]\pm\Delta_{m}\big\},
\\
\Delta_{m}^{2}=\{E+(N-m)[D_{1}+(N-1-m)G_{2}]\}^{2}\notag
\\
\phantom{\Delta_{m}^{2}=}
+4m\big\{(2N-1-m)A_{2}+B_{1}
+\big[3N^{2}-6N+2-3(N-1)m+m^{2}\big]F_{3}\big\}.
\end{gather}
\end{subequations}

These f\/indings imply that~-- at least in these cases~-- specif\/ic predictions on the actual behavior of
the solutions of the Newtonian~$N$-body problem~\eqref{EqMot} can be easily made.
In Section~\ref{section4} we identify in particular the cases in which this Newtonian~$N$-body
problem is \textit{isochronous} or \textit{asymptotically isochronous}, or an appropriate variant of it is
\textit{isochronous}; while the f\/indings described in this Section~\ref{section2} are proven in
the following Section~\ref{section3}.

\section[Derivation of the equations of motion of the Newtonian~$N$-body problem]{Derivation of the
equations of motion\\ of the Newtonian $\boldsymbol{N}$-body problem}
\label{section3}

Our task in this section is to obtain the equations of motion~\eqref{EqMot} characterizing the new~$N$-body
problem of \textit{goldfish} type.

The starting point is the \textit{linear} third-order PDE~\eqref{PDEpsi} satisf\/ied by the
polynomial~\eqref{psizt}.
We already saw in Section~\ref{section2} the implication of this evolution PDE for the
coef\/f\/icients $c_{m}(t)$ of its polynomial solution~\eqref{psizt}, leading to the identif\/ication~--
via~\eqref{psizt} and~\eqref{Solc}~-- of its solution by algebraic operations.
In this section we show that the fact that the polynomial~\eqref{psizt} satisf\/ies the PDE~\eqref{PDEpsi}
implies that its zeros~$z_{n}(t)$ indeed evolve according to the Newtonian equations of
motion~\eqref{EqMot}.

This is in fact an immediate consequence~-- via trivial, if somewhat cumbersome, algebra~-- of (some of)
the identities reported in Appendix~A of~\cite{C2008} and of the additional identities~\eqref{Iden3zp}
and~\eqref{Iden2zt} reported (and proven) in Appendix~\ref{appendixA} of this paper, see below.
Indeed these identities~-- valid for an \textit{arbitrary} time-dependent polynomial $\psi(z,t)$ of degree~$N$ in~$z$, see~\eqref{psizt})~-- allow to transform the \textit{linear} PDE~\eqref{PDEpsi} satisf\/ied by
the polynomial $\psi(z,t)$ into the system of \textit{nonlinear} ODEs~\eqref{EqMot} satisf\/ied by its
zeros $z_{n}(t)$.
Note that the fact that this outcome obtains is both a~\textit{consequence} and a~\textit{confirmation} of
the fact that the PDE~\eqref{PDEpsi} admits as its solution the polynomial $\psi(z,t)$, see~\eqref{psizt},
of degree~$N$ in~$z$, hence featuring~$N$ zeros~$z_{n}(t) $.

\section{Isochronous and asymptotically isochronous cases}
\label{section4}

In this section we identify cases in which the Newtonian~$N$-body problem~\eqref{EqMot} is
\textit{isochronous} or \textit{asymptotically isochronous}, or variants of it are \textit{isochronous}.

First of all we note, see~\eqref{Solc}, that if the $2N$ eigenvalues $ \lambda_{m}^{(\pm)}$ are
\textit{all imaginary} and read
\begin{gather}
\lambda_{m}^{(\pm)}=ir_{m}^{(\pm)}\omega,
\qquad
r_{m}^{(\pm)}\equiv\frac{p_{m}^{(\pm)}}{q_{m}^{(\pm)}},
\qquad
m=1,\dots,N,
\label{landaiso}
\end{gather}
-- with $\omega $ a~\textit{positive real} number, the~$N$ numbers $ q_{m}^{(\pm)}$ \textit{all positive
integers}, the~$N$ numbers $p_{m}^{(\pm)}$ \textit{all integers} (positive, negative or vanishing, with
$p_{m}^{(\pm)}$ and $q_{m}^{(\pm)}$ coprimes and the $2N$ \textit{real rational}
numbers $r_{m}^{(\pm)}\equiv p_{m}^{(\pm)}/q_{m}^{(\pm)}$ \textit{all different among themselves})~-- then the~$N$
coef\/f\/icients $c_{m}(t)$ evolve \textit{isochronously}, i.e.
\begin{subequations}
\begin{gather}
c_{m}(t+T)=c_{m}(t),
\qquad
m=1,\dots,N,
\end{gather}
with the period
\begin{gather}
T=\frac{2\pi q}{\omega}
\label{T}
\end{gather}
\end{subequations}
independent of the initial data.
Here of course $q$ is the \textit{minimum common multiple} of the $2N$ denominators of the $2N$
\textit{rational} numbers $r_{m}^{(\pm)}\equiv p_{m}^{(\pm)}/q_{m}^{(\pm)}$.
And it is plain that the same property of \textit{isochrony} is then shared by the~$N$ coordinates~$z_{n}(t)$, namely the Newtonian~$N$-body problem~\eqref{EqMot} is then \textit{isochronous} as well; with
the possibility that in some open regions of its phase space the periodicity only holds for a~period which
is a~(generally \textit{small}) \textit{integer multiple} of~$T$ due to the fact that some of the zeros of
the polynomial~$\psi(z,t)$~-- itself evolving \textit{isochronously} with period~$T$, see~\eqref{psizt}~--
might ``exchange their roles'' over the time evolution (for an analysis of this phenomenology, also
explaining the meaning of the assertion made above that the \textit{integer} \textit{multiple} in question
is generally \textit{small}, see~\cite{GS2005}).

Let us then focus on the two cases~-- as identif\/ied at the end of Section~\ref{section2},
see~\eqref{Res1} and~\eqref{Res2}~-- in which the $2N$ eigenvalues $\lambda_{m}^{(\pm)}$ can be
\textit{explicitly} obtained, see~\eqref{landas}.
It is then easy to identify the additional restrictions on the parameters which are \textit{necessary}~--
and also \textit{sufficient}, up to some minor additional restrictions to exclude the coincidence of
eigenvalues~-- to guarantee that the Newtonian~$N$-body problem~\eqref{EqMot} be \textit{isochronous} with
period~$T$.
They read:
\begin{subequations}
\begin{gather}
E=-2ir_{1}\omega,
\qquad
D_{1}=-2ir_{2}\omega,
\qquad
G_{2}=-2ir_{3}\omega,
\\
A_{2}=a_{2}\omega^{2},
\qquad
B_{1}=b_{1}\omega^{2},
\qquad
F_{3}=f_{3}\omega^{2},
\end{gather}
with
\begin{gather}
f_{3}=2r_{3}(s_{2}r_{4}-R_{2}),
\\
a_{2}=2(s_{1}s_{2}-1)R_{1}r_{3}-R_{2}^{2}+r_{4}^{2}-3(N-1)f_{3},
\\
b_{1}=2R_{1}(s_{1}r_{4}-R_{2})-(2N-1)a_{2}-\big(3N^{2}-6N+2\big)f_{3},
\\
R_{1}=r_{1}+N[r_{2}+(N-1)r_{3}],\qquad R_{2}=r_{2}+(2N-1)r_{3}.
\end{gather}
\end{subequations}
Here $r_{1}$, $r_{2}$, $r_{3}$, $r_{4}$ are 4 arbitrary \textit{rational} numbers and $s_{1}$ and $s_{2}$
are two \textit{arbitrary signs} ($+$~or~$-$).
Indeed with these assignments $\lambda_{m}^{(\pm)}$ clearly satisf\/ies the condition~\eqref{landaiso}
with
\begin{gather}
r_{m}^{(\pm)}=(1\pm s_{1})R_{1}-(R_{2}\pm r_{4})m+(1\pm1)r_{3}m^{2}.
\end{gather}

It is moreover plain that, if the $2N$ eigenvalues $\lambda_{m}^{(\pm)}$ satisfy, instead of the
condition~\eqref{landaiso}, the less restrictive condition
\begin{gather}
\lambda_{m}^{(\pm)}=\big(ir_{m}^{(\pm)}-\rho_{m}^{(\pm)}\big)\omega
\label{landarrho}
\end{gather}
-- with $\omega $ again a~\textit{positive real} number, the $2N$ numbers $ r_{m}^{(\pm)}$ again
\textit{all real} and \textit{rational} and the $2N$ numbers $\rho_{m}^{(\pm)}$ \textit{all real} and
\textit{nonnegative}, $\rho_{m}^{(\pm)}\geq 0$, with \textit{at least one of them vanishing}~-- then the
$N$ coef\/f\/icients $ c_{m}(t)$ are \textit{asymptotically isochronous},
\begin{subequations}
\begin{gather}
\underset{t\rightarrow\infty}\lim\big[c_{m}(t)-c_{m}^{\rm (asy)}(t)\big]=0,
\end{gather}
with $c_{m}^{\rm (asy)}(t)$ periodic,
\begin{gather}
c_{m}^{\rm (asy)}(t+T)=c_{m}^{\rm (asy)}(t).
\end{gather}
\end{subequations}
Here $T$ is given again by~\eqref{T}, but now with $q$ being the \textit{minimum common multiple} of the
denominators $q_{m}^{(\pm)}$'s of the rationals $r_{m}^{(\pm)}$'s associated with \textit{vanishing}
$\rho_{m}^{(\pm)}$'s, $\rho_{m}^{(\pm)}=0$, see~\eqref{landarrho} and~\eqref{landaiso}.
And it is again plain that the same property is then shared by the~$N$ coordinates~$ z_{n}(t)$, namely that
the Newtonian~$N$-body problem~\eqref{EqMot} is then \textit{asymptotically isochronous} as well (again,
with a~period which might be a, generally \textit{small}, \textit{integer multiple} of~$T$~\cite{GS2005}).

Next, less us investigate the cases in which~-- via a~well-known trick, see for instance
Section~2.1 (entitled ``The trick'') of~\cite{C2008}~-- \textit{isochronous} variants can be manufactured of the
Newtonian~$N$-body problem~\eqref{EqMot}.
To this end, it is convenient to re-write the equations of motion~\eqref{EqMot} via the \textit{formal}
replacement of dependent and independent variables~$z_{n}(t) \Rightarrow \zeta_{n}(\tau)$, so that the
equations of motion read as follows:
\begin{gather}
\zeta_{n}^{\prime\prime}+E\zeta_{n}^{\prime}=B_{0}+B_{1}\zeta_{n}-(N-1)[2A_{3}+3(N-2)F_{4}]\zeta_{n}^{2}
+(N-1)(N-2)G_{3}\zeta_{n}^{\prime}\zeta_{n}\notag
\\
\hphantom{\zeta_{n}^{\prime\prime}+E\zeta_{n}^{\prime}=}{}
+\sum_{\ell=1;\, \ell\neq n}^{N}\big\{(\zeta_{n}-\zeta_{\ell})^{-1}
\big[2\zeta_{n}^{\prime}\zeta_{\ell}^{\prime}+2\big(A_{0}+A_{1}\zeta_{n}+A_{2}\zeta_{n}^{2}+A_{3}\zeta_{n}^{3}\big)\notag
\\
\hphantom{\zeta_{n}^{\prime\prime}+E\zeta_{n}^{\prime}=}{}
-\big(\zeta_{n}^{\prime}+\zeta_{\ell}^{\prime}\big)(D_{0}+D_{1}\zeta_{n})-D_{2}\zeta_{n}(\zeta_{n}^{\prime}
\zeta_{\ell}+\zeta_{\ell}^{\prime}\zeta_{n})\big]\big\}\notag
\\
\hphantom{\zeta_{n}^{\prime\prime}+E\zeta_{n}^{\prime}=}{}
+\sum_{k,\ell=1;\,k\neq n,\, \ell\neq n,\, k\neq\ell}^{N}\bigg[\frac{3(F_{0}+F_{1}\zeta_{n}+F_{2}\zeta_{n}^{2}+F_{3}
\zeta_{n}^{3}+F_{4}\zeta_{n}^{4})}{(\zeta_{n}-\zeta_{\ell})(\zeta_{n}-\zeta_{k})}\label{EqMotzita}
\\
\hphantom{\zeta_{n}^{\prime\prime}+E\zeta_{n}^{\prime}=}{}
-\left(\frac{G_{0}+G_{1}\zeta_{n}+G_{2}\zeta_{n}^{2}+G_{3}\zeta_{n}^{3}}{\zeta_{n}-\zeta_{\ell}}\right)
\left(\frac{\zeta_{n}^{\prime}+\zeta_{k}^{\prime}}{\zeta_{n}-\zeta_{k}}+\frac{\zeta_{\ell}^{\prime}
+\zeta_{k}^{\prime}}{\zeta_{\ell}-\zeta_{k}}\right)\bigg],
\qquad
n=1,\dots,N.\notag
\end{gather}
Here and hereafter (in this section) appended primes denote dif\/ferentiation with respect to the variable~$\tau$.

\begin{subequations}
We now perform the following change of dependent and independent variables (``the trick''):
\begin{gather}
z_{n}(t)=\exp(i\alpha\omega t)\zeta_{n}(\tau),
\qquad
\tau\equiv\tau(t)=\frac{\exp(i\omega t)-1}{i\omega}.
\label{zzita}
\end{gather}
Here and below $\omega $ is a~positive constant, and $\alpha $ a~nonvanishing number that we reserve to
assign, see below.

It is then plain that there hold the following formulas:
\begin{gather}
\dot{z}_{n}-i\alpha\omega z_{n}=\exp[i(\alpha+1)\omega t]\zeta_{n}^{\prime},
\\
\ddot{z}_{n}-(2\alpha+1)i\omega\dot{z}_{n}-\alpha(\alpha+1)\omega^{2}z_{n}
=\exp[i(\alpha+2)\omega t]\zeta_{n}^{\prime\prime}.
\end{gather}
\end{subequations}
And via these formulas one can easily obtain the equations of motion implied for the dependent variables
$z_{n}(t)$ by the equations of motion~\eqref{EqMotzita} satisf\/ied by the variables $\zeta_{n}( \tau)$,
and thereby ascertain for which assignments of the parameter $\alpha$, and for which corresponding
restrictions on the 19 coupling constants featured by these equations of motion, the resulting equations of
motion satisf\/ied by the~$N$ coordinates $z_{n}(t)$ are \textit{autonomous}, i.e.\ they feature no
\textit{explicit} time-dependence.
We list below all these cases, on the understanding that \textit{all} the coupling constants which do
\textit{not} appear in these new equations of motion have been set to \textit{zero} (while those that do
appear are \textit{arbitrary}).

For $\alpha =-2$, these Newtonian equations of motion read
\begin{gather}
\ddot{z}_{n}+3i\omega\dot{z}_{n}-2\omega^{2}z_{n}=B_{0}+2\sum_{\ell=1;\, \ell\neq n}^{N}\frac{(\dot{z}_{n}
+2i\omega z_{n})(\dot{z}_{\ell}+2i\omega z_{\ell})}{z_{n}-z_{\ell}}\notag
\\
\hphantom{\ddot{z}_{n}+3i\omega\dot{z}_{n}-2\omega^{2}z_{n}=}
{} +\sum_{\ell,k=1;\, \ell\neq n,\, k\neq n,\, \ell\neq k}^{N}\frac{3F_{2}z_{n}^{2}}{(z_{n}-z_{\ell})(z_{n}-z_{k})}.
\end{gather}

For $\alpha =-1$, these Newtonian equations of motion read
\begin{gather}
\ddot{z}_{n}+i\omega\dot{z}_{n}=
\sum_{\ell=1;\, \ell\neq n}^{N}\frac{2(\dot{z}_{n}+i\omega z_{n})(\dot{z}_{\ell}+i\omega z_{\ell})+2A_{0}
-D_{0}[\dot{z}_{n}+\dot{z}_{\ell}+i\omega(z_{n}+z_{\ell})]}{z_{n}-z_{\ell}}\notag
\\
\phantom{\ddot{z}_{n}+i\omega\dot{z}_{n}=}
{} +\sum_{\ell,k=1;\, \ell\neq n,\, k\neq n,\, \ell\neq k}^{N}\bigg[\frac{3F_{1}z_{n}}{(z_{n}-z_{\ell})(z_{n}-z_{k})}\notag
\\
\phantom{\ddot{z}_{n}+i\omega\dot{z}_{n}=}
{} -\frac{G_{1}z_{n}}{z_{n}-z_{\ell}}\left(\frac{\dot{z}_{n}+\dot{z}_{k}+i\omega(z_{n}+z_{k})}{z_{n}-z_{k}}
+\frac{\dot{z}_{\ell}+\dot{z}_{k}+i\omega(z_{\ell}+z_{k})}{z_{\ell}-z_{k}}\right)\bigg].
\end{gather}

For $\alpha =-2/3$, these Newtonian equations of motion read
\begin{gather}
\ddot{z}_{n}+\frac{i\omega}{3}\dot{z}_{n}+\frac{2}{9}\omega^{2}z_{n}
=2\sum_{\ell=1;\, \ell\neq n}^{N}
\frac{(\dot{z}_{n}+2i\omega z_{n}/3)(\dot{z}_{\ell}+2i\omega z_{\ell}/3)}{z_{n}-z_{\ell}}\notag
\\
\phantom{\ddot{z}_{n}+\frac{i\omega}{3}\dot{z}_{n}+\frac{2}{9}\omega^{2}z_{n}=}{}
+\sum_{\ell,k=1;\, \ell\neq n,\, k\neq n,\, \ell\neq k}^{N}\frac{3F_{0}}{(z_{n}-z_{\ell})(z_{n}-z_{k})}.
\end{gather}

For $\alpha =-1/2$, these Newtonian equations of motion read
\begin{gather}
\ddot{z}_{n}+\frac{\omega^{2}}{4}z_{n}=2\sum_{\ell=1;\, \ell\neq n}^{N}
\frac{(\dot{z}_{n}+i\omega z_{n}/2)(\dot{z}_{\ell}+i\omega z_{\ell}/2)}{z_{n}-z_{\ell}}\notag
\\
\phantom{\ddot{z}_{n}+\frac{\omega^{2}}{4}z_{n}=}
{} -\sum_{\ell,k=1;\, \ell\neq n,\, k\neq n,\, \ell\neq k}^{N}\bigg[\frac{G_{0}}{z_{n}-z_{\ell}}\bigg(\frac{\dot{z}_{n}+\dot{z}
_{k}+i\omega(z_{n}+z_{k})/2}{z_{n}-z_{k}}\notag
\\
\phantom{\ddot{z}_{n}+\frac{\omega^{2}}{4}z_{n}=}
{} +\frac{\dot{z}_{\ell}+\dot{z}_{k}+i\omega(z_{\ell}+z_{k})/2}{z_{\ell}-z_{k}}\bigg)\bigg].
\end{gather}

For $\alpha =1$, these Newtonian equations of motion read
\begin{gather}
\ddot{z}_{n}-3i\omega\dot{z}_{n}-2\omega^{2}z_{n}=(N-1)(N-2)G_{3}(\dot{z}_{n}-i\omega z_{n})z_{n}\notag
\\
\phantom{\ddot{z}_{n}}{}
+\sum_{\ell=1;\, \ell\neq n}^{N}\big[(z_{n}-z_{\ell})^{-1}\big\{2(\dot{z}_{n}-i\omega z_{n})(\dot{z}_{\ell}
-i\omega z_{\ell})\notag
\\
\phantom{\ddot{z}_{n}}{}
-D_{2}z_{n}[(\dot{z}_{n}-i\omega z_{n})z_{\ell}+(\dot{z}_{\ell}-i\omega z_{\ell})z_{n}]\big\}\big]         \notag
\\
\phantom{\ddot{z}_{n}}{}
-\sum_{\ell,k=1;\, \ell\neq n,\, k\neq n,\, \ell\neq k}^{N}\bigg[\frac{G_{3}z_{n}^{3}}{z_{n}-z_{\ell}}\bigg(\frac{\dot{z}_{n}
+\dot{z}_{k}-i\omega(z_{n}+z_{k})}{z_{n}-z_{k}}
+\frac{\dot{z}_{\ell}+\dot{z}_{k}-i\omega(z_{\ell}+z_{k})}{z_{\ell}-z_{k}}\bigg)\bigg].
\end{gather}

For $\alpha =2$, these Newtonian equations of motion read
\begin{gather}
\ddot{z}_{n}-5i\omega\dot{z}_{n}-6\omega^{2}z_{n}=-(N-1)[2A_{3}+3(N-2)F_{4}]z_{n}^{2}
\\
\phantom{\ddot{z}_{n}}{}
+\sum_{\ell=1;\, \ell\neq n}^{N}\frac{2(\dot{z}_{n}-2i\omega z_{n})(\dot{z}_{\ell}-2i\omega z_{\ell})+2A_{3}
z_{n}^{3}}{z_{n}-z_{\ell}}
+\sum_{\ell,k=1;\, \ell\neq n,\, k\neq n,\, \ell\neq k}^{N}\frac{3F_{4}z_{n}^{4}}{(z_{n}-z_{\ell})(z_{n}-z_{k})}.\notag
\end{gather}

And it is plain that \textit{all} these Newtonian models are \textit{isochronous}: see~\eqref{zzita}, and
note that the solutions $\zeta_{n}(\tau)$ of the \textit{solvable}~$N$-body model~\eqref{EqMotzita} have at
most \textit{algebraic} singularities as functions of $ \tau $.

\section{Outlook}
\label{section5}

The search for more general \textit{solvable} models of goldf\/ish type should continue: they might still
yield interesting results.

Another development likely to yield interesting f\/indings is the investigation of the behavior, in the
inf\/initesimal neighborhood of its equilibria, of the~$N$-body model introduced above, especially in the
\textit{isochronous} cases.
This investigation might yield new \textit{Diophantine} f\/indings for the zeros of interesting polynomials.
We plan to pursue these results, which shall eventually be submitted to a~journal devoted to special
functions if they turn out to be suf\/f\/iciently interesting to justify their publication.

\appendix

\section{Appendix: identities involving the zeros of a~polynomial}
\label{appendixA}

In this Appendix we report (and then prove) several \textit{identities} for the time-dependent polynomial
$\psi \equiv \psi(z,t)$, see~\eqref{psizt}, of degree~$N$ in~$z$.
We of course use hereafter the notation introduced in Section~\ref{section2}, see
Notation~\ref{Notation2.1}, and in addition the following convenient shorthand notation:
as in~\cite{C2008} (see there equations~(A.4) and~(A.5))
\begin{subequations}
\label{DefD}
\begin{gather}
D\psi\Longleftrightarrow F_{n}(\underline{z},\underline{\dot{z}})
\end{gather}
-- with $D$ a~dif\/ferential operator acting on the independent variables~$z$ and~$t$ of the polynomial
$\psi(z,t)$, see~\eqref{psizt}~-- stands for the \textit{identity}
\begin{gather}
D\psi(z,t)=\psi(z,t)\sum_{n=1}^{N}[z-z_{n}(t)]^{-1}F_{n}(\underline{z},\underline{\dot{z}}).
\end{gather}
\end{subequations}
Below we often, for notational simplicity, omit to indicate explicitly the dependence on their arguments of
$\psi \equiv \psi(z,t)$, $z_{n}\equiv z_{n}(t)$ and $c_{m}(t)$ (see~\eqref{psizt}).

We now list the following identities, which complement those reported in Appendix~A of~\cite{C2008} (see in
particular the 2012 paperback version, where the formulas denoted in~\cite{C2008} as~(A.8k) and~(A.8l) are
corrected):
\begin{subequations}
\label{Iden3zp}
\begin{gather}
z^{p}\psi_{zzz}\ \Longleftrightarrow \ 3z_{n}^{p}\sum_{k,\ell=1;\, k\neq n,\, \ell\neq n,\, k\neq\ell}^{N}
\big[(z_{n}-z_{k})^{-1}(z_{n}-z_{\ell})^{-1}\big],
\qquad
p=0,1,2,
\label{Identzp}
\\
z^{3}\psi_{zzz}-N(N-1)(N-2)\psi
\ \Longleftrightarrow \ 3z_{n}^{3}\sum_{k,\ell=1;\, k\neq n,\, \ell\neq n,\, k\neq\ell}^{N}
\big[(z_{n}-z_{k})^{-1}(z_{n}-z_{\ell})^{-1}\big],
\label{Idenz3}
\\
z^{4}\psi_{zzz}+(N-1)(N-2)(3c_{1}-Nz)\psi\notag
\\
\qquad{}
\Longleftrightarrow \ 3z_{n}^{4}\sum_{k,\ell=1;\, k\neq n,\, \ell\neq n,\, k\neq\ell}^{N}
\big[(z_{n}-z_{k})^{-1}(z_{n}-z_{\ell})^{-1}\big];
\label{Idenz4}
\end{gather}
\end{subequations}
\vspace*{-6.25mm}
\begin{subequations}
\label{Iden2zt}
\begin{gather}
z^{p}\psi_{zzt}\ \Longleftrightarrow \ -\sum_{k,\ell=1;\, k\neq n,\, \ell\neq n,\, k\neq\ell}^{N}
\left[\left(\frac{z_{n}^{p}}{z_{n}-z_{k}}\right)
\left(\frac{\dot{z}_{n}+\dot{z}_{\ell}}{z_{n}-z_{\ell}}+\frac{\dot{z}_{k}+\dot{z}_{\ell}}
{z_{k}-z_{\ell}}\right)\right],
\qquad\!\!
p=0,1,2,\!\!\!\!
\label{Iden2ztp}
\\
z^{3}\psi_{zzt}-(N-1)(N-2)z\psi_{t} \ \Longleftrightarrow \ (N-1)(N-2)\dot{z}_{n}z_{n}
\notag
\\
\phantom{z^{3}\psi_{zzt}}{}
-\sum_{k,\ell=1;\,k\neq n,\, \ell\neq n,\, k\neq\ell}^{N}\left[\left(\frac{z_{n}^{3}}{z_{n}-z_{k}}\right)
\left(\frac{\dot{z}_{n}+\dot{z}_{\ell}}{z_{n}-z_{\ell}}+\frac{\dot{z}_{k}+\dot{z}_{\ell}}{z_{k}-z_{\ell}}\right)\right].
\label{Iden2zt3}
\end{gather}
\end{subequations}
Of course in the last two,~\eqref{Iden2ztp} and~\eqref{Iden2zt3}, superimposed dots indicate
$t$-dif\/ferentiations.
Also note that in the third,~\eqref{Idenz4}, of these identities, consistently with~\eqref{psizt}
\begin{gather}
c_{1}\equiv c_{1}(t)=-\sum_{n=1}^{N}[z_{n}(t)].
\label{c1}
\end{gather}
But remarkably~-- due to a~neat cancellation, see~\eqref{Idenz4} and equation~(A.6b) of~\cite{C2008}~-- this quantity does not appear in the equations of motion~\eqref{EqMot}.

Our task in this Appendix is to prove these \textit{identities}.
To perform these proofs it is convenient to introduce the following shorthand notation denoting sums over
(dummy) indices restricted to take \textit{different} values (among themselves):
\begin{gather}
\sum\nolimits_{nk}^{\prime}\equiv\sum_{n,k=1;\, n\neq k}^{N};
\qquad
\sum\nolimits_{nk\ell}^{\prime}\equiv\sum_{n,k,\ell=1;\, n\neq k,\, k\neq\ell,\, \ell\neq n}^{N}.
\label{NotSum}
\end{gather}

\begin{subequations}
We moreover introduce the following shorthand notation for the ``denominator'' $\den( z_{n},\!z_{k},\!z_{\ell})$,
\begin{gather}
\den(z_{n},z_{k},z_{\ell})\equiv(z_{n}-z_{k})(z_{n}-z_{\ell})(z_{k}-z_{\ell}),
\label{denden}
\end{gather}
which clearly has the property to be \textit{invariant} under the cyclic exchange $n\rightarrow
k\rightarrow \ell$ of the three indices $n$, $k$, $\ell$,
\begin{gather}
\den(z_{n},z_{k},z_{\ell})=\den(z_{k},z_{\ell},z_{n}),
\end{gather}
and to be instead \textit{antisymmetric} under the exchange of any two of its three arguments $z_{n}$,
$z_{k}$,~$z_{\ell}$,
\begin{gather}
\den(z_{n},z_{k},z_{\ell})=-\den(z_{k},z_{n},z_{\ell})=-\den(z_{\ell},z_{k},z_{n}
)=-\den(z_{n},z_{\ell},z_{k}).
\label{denanti}
\end{gather}
\end{subequations}
Likewise, we denote corresponding ``numerators'' as $\num(n,k,\ell)$
(i.e., $\num(z_{n},z_{k},z_{\ell})\equiv$ $\num(n$,                                
$k,\ell)$) and, whenever one of them is the sum of an
\textit{arbitrary} number of terms $\num_{j}(n,k,\ell)$, i.e.
\begin{gather}
\num(n,k,\ell)\equiv\sum_{j=0}\num_{j}(n,k,\ell),
\label{numer}
\end{gather}
we take advantage (if need be) of the following

\begin{Lemma}\label{LemmaA.1}
If  each  of the addends $\num_{j}(n,k,\ell)$ in the right-hand side of~\eqref{numer} is
 invariant  under the exchange of any two of the three indices $n$, $k$, $\ell$, namely if
 each  of the addends $\num_{j}(n,k,\ell)$ satisfies  at least one  of the following three
relations:
\begin{subequations}
\label{Symnum}
\begin{gather}
\num_{j}(n,k,\ell)=\num_{j}(k,n,\ell),
\end{gather}
or
\begin{gather}
\num_{j}(n,k,\ell)=\num_{j}(\ell,k,n),
\end{gather}
or
\begin{gather}
\num_{j}(n,k,\ell)=\num_{j}(n,\ell,k),
\end{gather}
\end{subequations}
then the triple sum $\sum\nolimits_{nk\ell}^{\prime}[ \num (n,k,\ell) /\den( z_{n},z_{k},z_{\ell})]$ vanishes:
\begin{gather}
\sum\nolimits_{nk\ell}^{\prime}\left[\frac{\num(n,k,\ell)}{\den(z_{n},z_{k},z_{\ell})}\right]
\equiv\sum\nolimits_{nk\ell}^{\prime}\bigg\{\sum_{j=0}
\left[\frac{\num_{j}(n,k,\ell)}{\den(z_{n},z_{k},z_{\ell})}\right]\bigg\}\notag
\\
\hphantom{\sum\nolimits_{nk\ell}^{\prime}\left[\frac{\num(n,k,\ell)}{\den(z_{n},z_{k},z_{\ell})}\right]}{}
=\sum_{j=0}\bigg\{\sum\nolimits_{nk\ell}^{\prime}
\left[\frac{\num_{j}(n,k,\ell)}{\den(z_{n},z_{k},z_{\ell})}\right]\bigg\}=0.
\label{Sumzero}
\end{gather}
\end{Lemma}

The validity of this assertion is an obvious consequence of the \textit{antisymmetry}~--
see~\eqref{denanti} and~\eqref{Symnum}~-- under the exchange of two appropriately chosen dummy indices in
the triple sums $ \sum\nolimits_{nk\ell}^{\prime}[ \num_{j}(n,k,\ell) /\den( z_{n},z_{k},z_{\ell})]$,
which therefore vanish (for all values of~$j$).

Next, let us report and prove the following identities, valid for any set of~$N$ arbitrary numbers~$z_{n}$
(for convenience, we always assume them to be \textit{all different among themselves}).
\begin{subequations}
\begin{gather}
\sum\nolimits_{nk}^{\prime}\left(\frac{z_{n}}{z_{n}-z_{k}}\right)=\frac{N(N-1)}{2},
\label{Id}
\\
\sum\nolimits_{nk\ell}^{\prime}\left(\frac{z_{k}}{z_{k}-z_{\ell}}\right)=\frac{N(N-1)(N-2)}{2},
\label{Sumkel}
\\
\sum\nolimits_{nk\ell}^{\prime}\left[\frac{z_{n}^{p}}{(z_{n}-z_{k})(z_{n}-z_{\ell})}\right]=0,
\qquad
p=0,1,
\label{Sumnp}
\\
\sum\nolimits_{nk\ell}^{\prime}\left[\frac{z_{n}^{2}}{(z_{n}-z_{k})(z_{n}-z_{\ell})}\right]=\frac{N(N-1)(N-2)}{3},
\label{Sumnkel}
\\
\sum\nolimits_{nk\ell}^{\prime}\left(\frac{z_{n}z_{k}}{z_{k}-z_{\ell}}\right)=-\frac{1}{2}(N-1)(N-2)c_{1},
\label{Sumnkelc1}
\\
\sum\nolimits_{nk\ell}^{\prime}\left(\frac{z_{n}^{2}}{z_{n}-z_{\ell}}\right)=-(N-1)(N-2)c_{1},
\label{Sumn2elc1}
\\
\sum\nolimits_{nk\ell}^{\prime}\left[\frac{z_{n}^{3}}{(z_{n}-z_{k})(z_{n}-z_{\ell})}\right]=-(N-1)(N-2)c_{1}.
\label{Sumzn3kel}
\end{gather}
\end{subequations}
In the last three formulas of course~$c_{1}$ is def\/ined by~\eqref{c1}.

The proof of the (well-known) identity~\eqref{Id} is trivial:
\begin{gather}
\sum\nolimits_{nk}^{\prime}\left(\frac{z_{n}}{z_{n}-z_{k}}\right)
=\frac{1}{2}\sum\nolimits_{nk}^{\prime}\left(\frac{z_{n}-z_{k}}{z_{n}-z_{k}}\right)=\frac{N(N-1)}{2}.
\end{gather}
The f\/irst step is justif\/ied by adding to the left-hand side of~\eqref{Id} what is obtained by the
exchange of the dummy indices $k$ and $\ell$ (which does not change the result) and dividing by $2$ (this
operation is often repeated below without describing it in as much detail as done here); the second step is
immediately implied by the f\/irst def\/inition~\eqref{NotSum}.

Then the proof of~\eqref{Sumkel} is no less trivial: it follows from~\eqref{Id} and the second
def\/ini\-tion~\eqref{NotSum}.

The proof of~\eqref{Sumnp} goes as follows.
For $p=0$ by adding to the right-hand side of this formula the sums obtained by performing
\textit{sequentially} two cyclic transformations of the three indices $n$, $k$, $\ell$ and by taking
advantage of the invariance property~\eqref{denden})~-- and by dividing the sum of the three equal sums
thereby obtained by~3~-- we clearly get
\begin{gather}
\sum\nolimits_{nk\ell}^{\prime}\left[\frac{1}{(z_{n}-z_{k})(z_{n}-z_{\ell})}\right]
=\frac{1}{3}\sum\nolimits_{nk\ell}^{\prime}\left[\frac{(z_{k}-z_{\ell})+(z_{\ell}-z_{n})+(z_{n}-z_{k})}
{\den(z_{n},z_{k},z_{\ell})}\right],
\end{gather}
and it is then plain that this quantity vanishes.
For $p$ $=1$ via~\eqref{denden}
\begin{gather}
\sum\nolimits_{nk\ell}^{\prime}\left[\frac{z_{n}}{(z_{n}-z_{k})(z_{n}-z_{\ell})}\right]
=\sum\nolimits_{nk\ell}^{\prime}\left[\frac{z_{n}z_{k}-z_{n}z_{\ell}}{\den(z_{n},z_{k},z_{\ell})}\right]
\end{gather}
and it is plain that this quantity vanishes thanks to  Lemma~\ref{LemmaA.1}, since in the right-hand side the
f\/irst term in the numerator is invariant under the exchange of dummy indices $n\leftrightarrow k$ and in
the second under the exchange $n\leftrightarrow \ell$.

The proof of~\eqref{Sumnkel} goes through the following steps:
\begin{gather}
\sum\nolimits_{nk\ell}^{\prime}\left[\frac{z_{n}^{2}}{(z_{n}-z_{k})(z_{n}-z_{\ell})}\right]
=\sum\nolimits_{nk\ell}^{\prime}\left[\frac{z_{n}^{2}}{(z_{n}-z_{k})(z_{n}-z_{\ell})}-\frac{z_{k}}{z_{k}
-z_{\ell}}\right]+\sum\nolimits_{nk\ell}^{\prime}\frac{z_{k}}{z_{k}-z_{\ell}}\notag
\\
\qquad{}
=-\sum\nolimits_{nk\ell}^{\prime}\left[\frac{z_{k}^{2}z_{\ell}-(z_{n}^{2}z_{k}+z_{k}^{2}z_{n}-z_{n}z_{k}
z_{\ell})}{\den(z_{n},z_{k},z_{\ell})}\right]+\frac{N(N-1)(N-2)}{2}\notag
\\
\qquad{}
=-\sum\nolimits_{nk\ell}^{\prime}\left[\frac{z_{k}^{2}z_{\ell}}{\den(z_{n},z_{k},z_{\ell})}
\right]+\frac{N(N-1)(N-2)}{2}\notag
\\
\qquad{}
=\sum\nolimits_{nk\ell}^{\prime}\left[\frac{z_{n}^{2}z_{\ell}}{\den(z_{n},z_{k},z_{\ell})}
\right]+\frac{N(N-1)(N-2)}{2}\notag
\\
\qquad{}
=-\frac{1}{2}\sum\nolimits_{nk\ell}^{\prime}\left[\frac{z_{n}^{2}}{(z_{n}-z_{k})(z_{n}-z_{\ell})}
\right]+\frac{N(N-1)(N-2)}{2},
\end{gather}
the f\/irst of which is quite trivial, the second is given by standard algebra together with the
def\/inition~\eqref{denden} and the formula~\eqref{Sumkel} just proven, the third of which is implied by
the formula~\eqref{Sumzero} (since $z_{n}^{2}z_{k}+z_{k}^{2}z_{n}-z_{n}z_{k}z_{\ell}$ is symmetrical under
the exchange of dummy indices $n\leftrightarrow k$), the fourth of which obtains via the exchange
$n\leftrightarrow k$ (see~\eqref{denanti}), and the last of which~-- which clearly
implies~\eqref{Sumnkel}~-- is clearly implied by the property~\eqref{denanti} and by the
def\/inition~\eqref{denden} of $\den( z_{n},z_{k},z_{\ell}) $.

The proof of~\eqref{Sumnkelc1} is again quite trivial:
\begin{gather}
\sum\nolimits_{nk\ell}^{\prime}\left(\frac{z_{n}z_{k}}{z_{k}-z_{\ell}}\right)=\frac{1}{2}
\sum\nolimits_{nk\ell}^{\prime}(z_{n})=-\frac{(N-1)(N-2)}{2}c_{1},
\end{gather}
the f\/irst step being implied by the antisymmetry of the summand under the exchange of the dummy indices
$k\leftrightarrow \ell$ and the second by the def\/initions of the symbol $\sum\nolimits_{nk\ell}^{\prime}$ (see~\eqref{NotSum}) and of~$c_{1}$ (see~\eqref{c1}).

The proof of~\eqref{Sumn2elc1} is analogous:
\begin{gather}
\sum\nolimits_{nk\ell}^{\prime}\left(\frac{z_{n}^{2}}{z_{n}-z_{\ell}}\right)=\frac{1}{2}
\sum\nolimits_{nk\ell}^{\prime}(z_{n}+z_{\ell})=\sum\nolimits_{nk\ell}^{\prime}(z_{n})
=-(N-1)(N-2)c_{1},
\end{gather}
with the f\/irst step justif\/ied by the replacement $z_{n}^{2}\rightarrow ( z_{n}^{2}-z_{\ell }^{2}) /2$
due to the antisymmetry of the summand under the exchange of dummy indices $n\leftrightarrow \ell$, the
second step justif\/ied by the symmetry of the summand, and the third step justif\/ied as just above.

Finally, the proof of~\eqref{Sumzn3kel} is analogous:
\begin{gather}
\sum\nolimits_{nk\ell}^{\prime}\left[\frac{z_{n}^{3}}{(z_{n}-z_{k})(z_{n}-z_{\ell})}\right]=\sum\nolimits_{nk\ell}
^{\prime}\left[\frac{z_{n}^{3}}{(z_{n}-z_{k})(z_{n}-z_{\ell})}-\frac{z_{n}^{2}}{z_{k}-z_{\ell}}\right]\notag
\\
\qquad{}
=-\sum\nolimits_{nk\ell}^{\prime}\left[\frac{z_{n}^{2}(z_{n}-z_{k})}{(z_{n}-z_{\ell})(z_{k}-z_{\ell})}
\right]=-\frac{1}{2}\sum\nolimits_{nk\ell}^{\prime}\left[\frac{(z_{n}^{2}-z_{k}^{2})(z_{n}-z_{k})}{(z_{n}-z_{\ell}
)(z_{k}-z_{\ell})}\right]\notag
\\
\qquad{}
=\frac{1}{2}\sum\nolimits_{nk\ell}^{\prime}
\left(\frac{z_{n}^{2}-z_{k}^{2}}{z_{n}-z_{\ell}}-\frac{z_{n}^{2}
-z_{k}^{2}}{z_{k}-z_{\ell}}\right)
=\frac{1}{2}\sum\nolimits_{nk\ell}^{\prime}
\left(\frac{z_{n}^{2}}{z_{n}-z_{\ell}}+\frac{z_{k}^{2}}{z_{k}-z_{\ell}}\right)\notag
\\
\qquad{}
=\sum\nolimits_{nk\ell}^{\prime}\left(\frac{z_{n}^{2}}{z_{n}-z_{\ell}}\right)=-(N-1)(N-2)c_{1}.
\end{gather}
Here the f\/irst step is justif\/ied by the vanishing of (the sum over) the added term (due to the
antisymmetry of the summand under the exchange of dummy indices $k\leftrightarrow \ell$), the second step
follows by trivial algebra, the third step is justif\/ied by the symmetry of the summand under the exchange
of dummy indices $n\leftrightarrow k$, the fourth step is justif\/ied by the identity
\begin{gather}
-\frac{z_{n}-z_{k}}{(z_{n}-z_{\ell})(z_{k}-z_{\ell})}=\frac{1}{z_{n}-z_{\ell}}-\frac{1}{z_{k}-z_{\ell}},
\end{gather}
the f\/ifth step by the elimination of two addends antisymmetric under the exchanges of dummy indices
$n\leftrightarrow \ell$ respectively $ k\leftrightarrow \ell$, the sixth step by the symmetry of the
summand under the exchange of dummy indices $n\leftrightarrow \ell$, and the last step by~\eqref{Sumn2elc1}.

Now we can f\/inally proceed and prove the formulas~\eqref{Iden3zp} and~\eqref{Iden2zt}.

We start from reporting equations~(A.1) (which coincides with~\eqref{psizt}), (A.2), (A.3), (A.8a) and (A.9a) of~\cite{C2008}:
\begin{gather}\label{psi}
\psi=\prod\limits_{n=1}^{N}[z-z_{n}]=z^{N}+\sum_{m=1}^{N}c_{m}z^{N-m}=\sum_{m=0}^{N}c_{m}z^{N-m},
\qquad
c_{0}=1,
\\
\psi_{z}=\psi\sum_{n=1}^{N}(z-z_{n})^{-1},
\label{psiz}
\\
\psi_{t}=-\psi\sum_{n=1}^{N}(z-z_{n})^{-1}\dot{z}_{n},
\label{psit}
\\
\psi_{zz}=2\psi\sum_{n=1}^{N}\Bigg[(z-z_{n})^{-1}\sum_{\ell=1;\,\ell\neq n}^{N}(z_{n}-z_{\ell})^{-1}\Bigg],
\label{psizz}
\\
\psi_{zt}=-\psi\sum_{n=1}^{N}\bigg\{(z-z_{n})^{-1}\sum_{\ell=1;\, \ell\neq n}^{N}\big[(\dot{z}_{n}+\dot{z}_{\ell}
)(z_{n}-z_{\ell})^{-1}\big]\bigg\}.
\label{Psizt}
\end{gather}
The last two equations correspond of course to (A8.a) and (A.9a) via the convention def\/ining the symbol
$\Longleftrightarrow $, see~\eqref{DefD}.

Partial dif\/ferentiation of~\eqref{psizz} with respect to~$z$ gives
\begin{subequations}
\begin{gather}
\psi_{zzz}=2\psi_{z}\sum_{n=1}^{N}(z-z_{n})^{-1}\!
\sum_{\ell=1;\,\ell\neq n}^{N}\!(z_{n}-z_{\ell})^{-1}-2\psi
\sum_{n=1}^{N}(z-z_{n})^{-2}\!
\sum_{\ell=1;\,\ell\neq n}^{N}(z_{n}-z_{\ell})^{-1}.\!\!
\end{gather}

Via~\eqref{psiz} this becomes (after a~convenient cancellation and change of dummy index from $n$ to $k$)
\begin{gather}
\psi_{zzz}=2\psi\sum_{n=1}^{N}
\Bigg[(z-z_{n})^{-1}\sum_{k=1;\,k\neq n}^{N}(z-z_{k})^{-1}
\sum_{\ell=1;\,\ell\neq n}^{N}(z_{n}-z_{\ell})(z_{n}-z_{\ell})\Bigg].
\end{gather}
\end{subequations}

We then use the identity
\begin{gather}\label{Iden}
(z-z_{n})^{-1}(z-z_{k})^{-1}=(z_{n}-z_{k})^{-1}\big[(z-z_{n})^{-1}-(z-z_{k})^{-1}\big],
\end{gather}
getting thereby
\begin{subequations}
\begin{gather}
\psi_{zzz}=2\psi\sum_{n,k=1;\, k\neq n}^{N}\Bigg\{\big[(z-z_{n})^{-1}-(z-z_{k})^{-1}\big]
(z_{n}-z_{k})^{-1}\sum_{\ell=1;\, \ell\neq n}^{N}(z_{n}-z_{\ell})^{-1}\Bigg\}.
\end{gather}
We then exchange the dummy indices $n$ and $k$ in the second of the two sums over these indices, getting
thereby
\begin{gather}
\psi_{zzz}=2\psi\!\!\sum_{n,k=1;\,k\neq n}^{N}\!\!\Bigg\{\frac{1}{(z-z_{n})(z_{n}-z_{k})}
\Bigg[\sum_{\ell=1;\,\ell\neq n}^{N}\left(\frac{1}{z_{n}-z_{\ell}}\right)
+\!\!\sum_{\ell=1;\,\ell\neq k}^{N}\!\!
\left(\frac{1}{z_{k}-z_{\ell}}\right)\Bigg]\Bigg\},\!\!
\end{gather}
which can also be written as follows:
\begin{gather}
\psi_{zzz}=2\psi\sum_{n=1}^{N}\Bigg\{\left(\frac{1}{z-z_{n}}\right)
\sum_{k,\ell=1;\, k\neq n,\, \ell\neq n,\, k\neq\ell}^{N}\bigg[\left(\frac{1}{z_{n}-z_{k}}\right)
\bigg(\frac{1}{z_{n}-z_{\ell}}
+\frac{1}{z_{k}-z_{\ell}}\bigg)\bigg]\Bigg\}.\!\!\!
\end{gather}
\end{subequations}
This implies (using the def\/inition of the symbols $\Leftrightarrow $ and $ \sum\nolimits_{nk\ell}^{\prime}$, see above)
\begin{subequations}
\begin{gather}
\psi_{zzz}\ \Longleftrightarrow\ 2\sum_{k,\ell=1;\, k\neq n,\, \ell\neq n,\, k\neq\ell}^{N}\big\{(z_{n}-z_{k})^{-1}
\big[(z_{n}-z_{\ell})^{-1}+(z_{k}-z_{\ell})^{-1}\big]\big\},
\\
\psi_{zzz}\ \Longleftrightarrow\ 2\sum\nolimits_{nk\ell}^{\prime}
\big[(z_{n}+z_{k}-2z_{\ell})(z_{k}-z_{\ell})^{-1}
(z_{n}-z_{k})^{-1}(z_{n}-z_{\ell})^{-1}\big];
\end{gather}
\end{subequations}
and f\/inally, taking advantage of the fact that the part of the summand antisymmetric under the exchange
of the two dummy indices $k$ and $\ell$ can be eliminated~-- so that $( z_{n}+z_{k}-2z_{\ell}) /(
z_{k}-z_{\ell})$ $\equiv 3/2+( 2z_{n}-z_{k}-z_{\ell}) / [ 2( z_{k}-z_{\ell}) ] $ can be replaced by
$3/2$~-- one gets~\eqref{Identzp} with $p=0$, which is thereby proven.
This is the f\/irst of the new formulas analogous to those reported in Appendix~A of~\cite {C2008}.

To proceed it is convenient to write in longhand the formula we just proved:
\begin{gather}
\psi_{zzz}=\psi\sum_{n=1}^{N}
\Bigg\{3(z-z_{n})^{-1}\sum_{k,\ell=1;\,k\neq n,\, \ell\neq n,\, k\neq\ell}^{N}
\big[(z_{n}-z_{k})^{-1}(z_{n}-z_{\ell})^{-1}\big]\Bigg\}.
\end{gather}
Multiplication by $z^{p}$ then yields
\begin{gather}
z^{p}\psi_{zzz}=\psi\sum_{n=1}^{N}\Bigg\{3 z^{p}(z-z_{n})^{-1}
 \sum_{k,\ell=1;\, k\neq n,\, \ell\neq n,\, k\neq\ell}^{N}
\big[(z_{n}-z_{k})^{-1}(z_{n}-z_{\ell})^{-1}\big]\Bigg\}.
\label{zpsizzz}
\end{gather}
By replacing the $z^{p}$ in the numerator with $z_{n}^{p}+( z^{p}-z_{n}^{p})$ we get
\begin{gather}
z^{p}\psi_{zzz}=
\psi\sum_{n=1}^{N}\Bigg\{\frac{3z_{n}^{p}}{z-z_{n}}
\sum_{k,\ell=1;\,k\neq n,\,\ell\neq n,\, k\neq\ell}^{N}
\big[(z_{n}-z_{k})^{-1}(z_{n}-z_{\ell})^{-1}\big]\Bigg\}\notag
\\
\phantom{z^{p}\psi_{zzz}=}
{}+3\psi\sum\nolimits_{nk\ell}^{\prime}
\left[\frac{z^{p}-z_{n}^{p}}{z-z_{n}}(z_{n}-z_{k})^{-1}(z_{n}-z_{\ell})^{-1}\right]
\label{psizzzzp}
\end{gather}
(see~\eqref{NotSum}).
It is then plain that, for $p=1$ and $p=2$, the formula~\eqref{Sumnp} implies that the last sum in this
equation vanishes.
Hence~\eqref{Identzp} is now proven also for $p=1$ and $p=2$.

To prove~\eqref{Idenz3} we set $p=3$ in~\eqref{psizzzzp}, which then reads
\begin{gather}
z^{3}\psi_{zzz}=
\psi\sum_{n=1}^{N}\Bigg\{\frac{3z_{n}^{3}}{z-z_{n}}\sum_{k,\ell=1;\, k\neq n,\, \ell\neq n,\, k\neq\ell}^{N}
\big[(z_{n}-z_{k})^{-1}(z_{n}-z_{\ell})^{-1}\big]\Bigg\}\notag
\\
\phantom{z^{3}\psi_{zzz}=}
+3\psi\sum\nolimits_{nk\ell}^{\prime}\big[(z^{2}+zz_{n}+z_{n}^{2})(z_{n}-z_{k})^{-1}(z_{n}-z_{\ell})^{-1}\big]
\end{gather}
(via the identity $z^{3}-z_{n}^{3}=(z-z_{n}) ( z^{2}+zz_{n}+z_{n}^{2}) $).
And it is then plain that this formula, via~\eqref{Sumnp} and~\eqref{Sumnkel}, yields~\eqref{Idenz3}, which
is thereby proven.

Likewise, to prove~\eqref{Idenz4} we set $p=4$ in~\eqref{psizzzzp}, which then reads
\begin{gather}
z^{4}\psi_{zzz}=
\psi\sum_{n=1}^{N}\bigg\{3(z-z_{n})^{-1}z_{n}^{4}\sum_{k,\ell=1;\,k\neq n,\, \ell\neq n,\, k\neq\ell}^{N}
\big[(z_{n}-z_{k})^{-1}(z_{n}-z_{\ell})^{-1}\big]\bigg\}\notag
\\
\phantom{z^{4}\psi_{zzz}=}
{}+3\psi\sum\nolimits_{nk\ell}^{\prime}
\big[\big(z^{3}+z^{2}z_{n}+zz_{n}^{2}+z_{n}^{3}\big)(z_{n}-z_{k})^{-1}(z_{n}-z_{\ell})^{-1}\big].
\end{gather}
Then it is easily seen that, via~\eqref{Sumnp},~\eqref{Sumnkel} and~\eqref{Sumzn3kel}, this equation
yields~\eqref{Idenz4}, which is thereby proven.

Next, to prove~\eqref{Iden2ztp} (to begin with, with $p=0$), we~$z$-dif\/ferentiate~\eqref{Psizt} getting
thereby
\begin{gather}
\psi_{zzt}=-\psi_{z}\sum_{n=1}^{N}
\bigg[(z-z_{n})^{-1}\sum_{\ell=1;\,\ell\neq n}^{N}\left(\frac{\dot{z}_{n}+\dot{z}
_{\ell}}{z_{n}-z_{\ell}}\right)\bigg]
+\psi\sum_{n=1}^{N}
\bigg[(z-z_{n})^{-2}\sum_{k=1;\,k\neq n}^{N}\left(\frac{\dot{z}_{n}+\dot{z}_{k}}{z_{n}-z_{k}}\right)\bigg]\notag
\\
\phantom{\psi_{zzt}}{}
=-\psi\sum_{k=1}^{N}\bigg\{(z-z_{k})^{-1}\sum_{n=1;\, n\neq k}^{N}\bigg[(z-z_{n})^{-1}\sum_{\ell=1;\, \ell\neq n}^{N}
\left(\frac{\dot{z}_{n}+\dot{z}_{\ell}}{z_{n}-z_{\ell}}\right)\bigg]\bigg\}\notag
\\
\phantom{\psi_{zzt}}{}
=-\psi\sum\nolimits_{nk\ell}^{\prime}\bigg\{\bigg[(z-z_{k})^{-1}-(z-z_{n})^{-1}\bigg](z_{k}-z_{n})^{-1}
\left(\frac{\dot{z}_{n}+\dot{z}_{\ell}}{z_{n}-z_{\ell}}\right)\bigg\}\notag
\\
\phantom{\psi_{zzt}}{}
=-\psi\sum_{n=1}^{N}\frac{1}{z-z_{n}}
\Bigg\{\sum_{k,\ell=1;\,k\neq n,\,\ell\neq n,\,\ell\neq k}^{k}
\bigg[\frac{1}{z_{n}-z_{k}}
\left(\frac{\dot{z}_{n}+\dot{z}_{\ell}}{z_{n}-z_{\ell}}
+\frac{\dot{z}_{k}+\dot{z}_{\ell}}{z_{k}-z_{\ell}}\right)\bigg]\Bigg\}.
\end{gather}
The f\/irst equality corresponds clearly to the dif\/ferentiation of~\eqref{Psizt}; the second obtains
via~\eqref{psiz} (also taking account of the cancellation occurring for $n=k$); the third equality obtains
via the identity
\begin{gather}
(z-z_{k})^{-1}(z-z_{n})^{-1}=\big[(z-z_{k})^{-1}-(z-z_{n})^{-1}\big](z_{k}-z_{n})^{-1};
\end{gather}
and the fourth equality obtains via the exchange of dummy indices $ k\leftrightarrow n$ in the sum
containing the term $(z-z_{k})^{-1}$ (and note the cancellations of the term with $\ell =n$ in the
resulting sum, with the term with $\ell =k$ in the second sum, justifying the exclusion of the 3 addends
with $k=n$, $\ell =n$ and $\ell =k$ in the sum in the last formula).
The f\/inal result corresponds~-- of course, via the notation~\eqref{DefD}~-- to~\eqref{Iden2ztp} with
$p=0$, which is thereby proven.

To prove~\eqref{Iden2ztp} with $p=1$ and $p=2$ we start from the longhand version of the equation we just
proved, multiplied by $z^{p}\equiv z_{n}^{p}+( z^{p}-z_{n}^{p})$:
\begin{gather}
z^{p}\psi_{zzt}=
-\psi\sum\nolimits_{nk\ell}^{\prime}\bigg\{\frac{z_{n}^{p}}{z-z_{n}}\bigg[(z_{n}-z_{k})^{-1}
\left(\frac{\dot{z}_{n}+\dot{z}_{\ell}}{z_{n}-z_{\ell}}
+\frac{\dot{z}_{k}+\dot{z}_{\ell}}{z_{k}-z_{\ell}}\right)\bigg]\bigg\}\notag
\\
\phantom{z^{p}\psi_{zzt}=}
-\psi\sum\nolimits_{nk\ell}^{\prime}\bigg\{\left(\frac{z^{p}-z_{n}^{p}}{z-z_{n}}\right)\bigg[(z_{n}-z_{k})^{-1}
\left(\frac{\dot{z}_{n}+\dot{z}_{\ell}}{z_{n}-z_{\ell}}
+\frac{\dot{z}_{k}+\dot{z}_{\ell}}{z_{k}-z_{\ell}}\right)\bigg]\bigg\}.
\end{gather}
From this formula it is plain that~\eqref{Iden2ztp} is proven also for $p=1$ and $p=2$ if, for these two
values of $p$, there holds the \textit{identity}
\begin{gather}
\sigma(p;\underline{z},\underline{\dot{z}})\equiv\sum\nolimits_{nk\ell}^{\prime}
\bigg[\left(\frac{z^{p}-z_{n}^{p}}{z-z_{n}}\right)\left(\frac{1}{z_{n}-z_{k}}\right)
\left(\frac{\dot{z}_{n}+\dot{z}_{\ell}}{z_{n}-z_{\ell}}
+\frac{\dot{z}_{k}+\dot{z}_{\ell}}{z_{k}-z_{\ell}}\right)\bigg]=0.
\label{sigmap}
\end{gather}
Note that we have now employed the convenient notation~\eqref{NotSum}.

Indeed, for $p=1$,
\begin{gather}
\sigma(1;\underline{z},\underline{\dot{z}})=\sum\nolimits_{nk\ell}^{\prime}
\left[\left(\frac{1}{z_{n}-z_{k}}\right)
\left(\frac{\dot{z}_{n}+\dot{z}_{\ell}}{z_{n}-z_{\ell}}+\frac{\dot{z}_{k}+\dot{z}_{\ell}}{z_{k}-z_{\ell}}\right)
\right]\notag
\\
\phantom{\sigma(1;\underline{z},\underline{\dot{z}})}{}
=\sum\nolimits_{nk\ell}^{\prime}
\left[\frac{\dot{z}_{n}z_{k}+\dot{z}_{k}z_{n}+\dot{z}_{\ell}(z_{n}+z_{k}
)-(\dot{z}_{n}+\dot{z}_{k}+2\dot{z}_{\ell})z_{\ell}}{\den(z_{n},z_{k},z_{\ell})}\right].
\end{gather}
The second equality obtains via trivial algebra and the def\/inition~\eqref{denden} of $\den(
z_{n},z_{k},z_{\ell})$; and it clearly implies that $\sigma ( 1;\underline{z},\underline{\dot{z}})$
vanishes via  Lemma~\ref{LemmaA.1}, since the numerator of the summand in the sum in the right-hand side of
the second equality is clearly invariant under the exchange of dummy indices $n\leftrightarrow k$.

Likewise, for $p=2$,
\begin{gather}
\sigma(2;\underline{z},\underline{\dot{z}})=\sum\nolimits_{nk\ell}^{\prime}
\left[\left(\frac{z+z_{n}}{z_{n}-z_{k}}\right)
\left(\frac{\dot{z}_{n}+\dot{z}_{\ell}}{z_{n}-z_{\ell}}+\frac{\dot{z}_{k}+\dot{z}_{\ell}}{z_{k}
-z_{\ell}}\right)\right]
=z\sigma(1;\underline{z},\underline{\dot{z}})
\notag
\\
\phantom{\sigma(2;\underline{z},\underline{\dot{z}})=}{}
+\sum\nolimits_{nk\ell}^{\prime}\left\{\left[\frac{z_{n}[\dot{z}_{n}z_{k}+\dot{z}_{k}z_{n}+\dot{z}_{\ell}(z_{n}
+z_{k})-(\dot{z}_{n}+\dot{z}_{k}+2\dot{z}_{\ell})z_{\ell}]}{\den(z_{n},z_{k},z_{\ell})}\right]\right\}
\notag
\\
\phantom{\sigma(2;\underline{z},\underline{\dot{z}})}{}
=\sum\nolimits_{nk\ell}^{\prime}\left\{\left[\frac{z_{n}[\dot{z}_{n}z_{k}+\dot{z}_{k}z_{n}+\dot{z}_{\ell}(z_{n}
+z_{k})-(\dot{z}_{n}+\dot{z}_{k}+2\dot{z}_{\ell})z_{\ell}]}{\den(z_{n},z_{k},z_{\ell})}\right]\right\}
\notag
\\
\phantom{\sigma(2;\underline{z},\underline{\dot{z}})}{}
=\sum\nolimits_{nk\ell}^{\prime}\big\{\big[\den(z_{n},z_{k},z_{\ell})]^{-1}\times
\notag
\\
\phantom{\sigma(2;\underline{z},\underline{\dot{z}})=}{}
\times
\big[(\dot{z}_{k}+\dot{z}_{\ell})z_{n}^{2}-(\dot{z}_{n}+\dot{z}_{\ell})z_{n}z_{\ell}+\dot{z}_{n}z_{n}z_{k}
+\dot{z}_{\ell}z_{k}z_{n}-(\dot{z}_{k}+\dot{z}_{\ell})z_{n}z_{\ell}\big]\big\}
\notag
\\
\phantom{\sigma(2;\underline{z},\underline{\dot{z}})}
=\sum\nolimits_{nk\ell}^{\prime}\left\{\left[\frac{\dot{z}_{n}z_{n}z_{k}+\dot{z}_{\ell}z_{k}z_{n}-\dot{z}_{k}
z_{n}z_{\ell}-\dot{z}_{\ell}z_{n}z_{\ell}}{\den(z_{n},z_{k},z_{\ell})}\right]\right\}
\notag
\\
\phantom{\sigma(2;\underline{z},\underline{\dot{z}})}{}
=\sum\nolimits_{nk\ell}^{\prime}\left\{\left[\frac{\dot{z}_{n}z_{n}(z_{k}+z_{\ell})+2\dot{z}_{\ell}z_{k}z_{n}}
{\den(z_{n},z_{k},z_{\ell})}\right]\right\}=0.
\end{gather}
The f\/irst equality is implied by the def\/inition~\eqref{sigmap} with $p=2$ and the identity
$z^{2}-z_{n}^{2}=( z+z_{n}) (z-z_{n}) $; the second equality is a~consequence of the vanishing
of $\sigma(1;\underline{z},\underline{\dot{z}})$ proven above; the third equality is obtained via trivial algebra;
the fourth equality obtains~-- thanks to  Lemma~\ref{LemmaA.1}~-- because the f\/irst two addends in the
numerator in the right-hand side are invariant under the exchanges of dummy indices $k\leftrightarrow \ell$
respectively $n\leftrightarrow \ell$; the f\/ifth equality obtains by performing on the fourth term in the
numerator the exchange of dummy indices $n\leftrightarrow \ell$ (entailing a~change of sign, since the
denominator changes sign under this exchange of indices, see~\eqref{denanti}) and on the third term the
exchange of dummy indices $k\leftrightarrow \ell$ (entailing likewise a~change of sign,
see~\eqref{denanti}).
Finally the last equality is implied by  Lemma~\ref{LemmaA.1}, since the f\/irst term in the numerator is now
invariant under the exchange of dummy indices $k\leftrightarrow \ell$ and the second under the exchange $
n\leftrightarrow k$.

Finally, let us prove~\eqref{Iden2zt3}.
Now we start from the identity
\begin{gather}
z^{2}\psi_{zzt}-(N-1)(N-2)\psi_{t}=(N-1)(N-2)\psi\sum_{n=1}^{N}\left(\frac{\dot{z}_{n}}{z-z_{n}}\right)
\notag
\\
\phantom{z^{2}\psi_{zzt}}{}
-\psi\sum\nolimits_{nk\ell}^{\prime}\bigg\{\frac{z_{n}^{2}}{z-z_{n}}\left[(z_{n}-z_{k})^{-1}
\left(\frac{\dot{z}_{n}+\dot{z}_{\ell}}{z_{n}-z_{\ell}}
+\frac{\dot{z}_{k}+\dot{z}_{\ell}}{z_{k}-z_{\ell}}\right)\right]\bigg\}
\notag
\\
\qquad{}
=-\psi\sum\nolimits_{nk\ell}^{\prime}\bigg\{\frac{1}{z-z_{n}}\bigg[\frac{z_{n}^{2}}{z_{n}-z_{k}}
\left(\frac{\dot{z}_{n}+\dot{z}_{\ell}}{z_{n}-z_{\ell}}
+\frac{\dot{z}_{k}+\dot{z}_{\ell}}{z_{k}-z_{\ell}}\right)-\dot{z}_{n}\bigg]\bigg\}.
\end{gather}
This identity is justif\/ied by~\eqref{psit} and by (the longhand version of) the identity we just
proved, \eqref{Iden2ztp}.
To write the second equality we also used the trivial observation that the very def\/inition of the symbol
$ \sum\nolimits_{nk\ell}^{\prime}$, see~\eqref{NotSum}, implies the identity $\sum\nolimits_{nk\ell
}^{\prime}f(n) =( N-1)(N-2)\sum\limits_{n=1}^{N}f(n)$ for any function $f(n)$ (and we will feel free
to use this identity again below).

We now multiply this identity by $z\equiv (z-z_{n}) +z_{n}$, and we thereby obtain
\begin{subequations}
\begin{gather}
z^{2}\psi_{zzt}-(N-1)(N-2)\psi_{t}=
-\tilde{\sigma}(3;\underline{z},\underline{\dot{z}})\psi
+\psi\sum_{n=1}^{N}\bigg[\left(\frac{1}{z-z_{n}}\right)\bigg\{(N-1)(N-2)\dot{z}_{n}z_{n}
\notag
\\
\hphantom{z^{2}\psi_{zzt}}{}
-\sum_{k,\ell=1;\,k\neq n,\,\ell\neq n,\,\ell\neq k}\bigg[\frac{z_{n}^{3}}{z-z_{n}}
\left(\frac{\dot{z}_{n}+\dot{z}_{\ell}}{z_{n}-z_{\ell}}
+\frac{\dot{z}_{k}+\dot{z}_{\ell}}{z_{k}-z_{\ell}}\right)\bigg]\bigg\}\bigg],
\end{gather}
with
\begin{gather}\label{sigma3}
\tilde{\sigma}(3;\underline{z},\underline{\dot{z}})\equiv
\sum\nolimits_{nk\ell}^{\prime}
\bigg[\frac{z_{n}^{2}}
{z_{n}-z_{k}}\left(\frac{\dot{z}_{n}+\dot{z}_{\ell}}{z_{n}-z_{\ell}}+\frac{\dot{z}_{k}+\dot{z}_{\ell}}{z_{k}
-z_{\ell}}\right)-\dot{z}_{n}\bigg].
\end{gather}
\end{subequations}

It is now plain that~\eqref{Iden2zt3} is proven if we show that
$\tilde{\sigma}(3;\underline{z},\underline{\dot{z}})$ vanishes,
$\tilde{\sigma}(3;\underline{z},\underline{\dot{z}})=0$.
To prove this we make the following steps:
\begin{gather}
\tilde{\sigma}(3;\underline{z},\underline{\dot{z}})=
\sum\nolimits_{nk\ell}^{\prime}\big([\den(z_{n},z_{k},z_{\ell})]^{-1}
\big\{z_{n}^{2}[(\dot{z}_{n}+\dot{z}_{\ell})(z_{k}-z_{\ell})+(\dot{z}_{k}+\dot{z}_{\ell})(z_{n}
-z_{\ell})]\notag
\\
\phantom{\tilde{\sigma}(3;\underline{z},\underline{\dot{z}})=}{}
-\dot{z}_{n}(z_{n}-z_{k})(z_{n}-z_{\ell})(z_{k}-z_{\ell})\big\}\big)\notag
\\
\phantom{\tilde{\sigma}(3;\underline{z},\underline{\dot{z}})}{}
=\sum\nolimits_{nk\ell}^{\prime}\big([\den(z_{n},z_{k},z_{\ell})]^{-1}
\big\{\dot{z}_{n}(z_{k}-z_{\ell})\big[z_{n}^{2}-(z_{n}-z_{k})(z_{n}-z_{\ell})\big]\notag
\\
\phantom{\tilde{\sigma}(3;\underline{z},\underline{\dot{z}})=}{}
+\dot{z}_{k}z_{n}^{2}(z_{n}-z_{\ell})+\dot{z}_{\ell}z_{n}^{2}(z_{n}+z_{k}-2z_{\ell})\big\}\big)\notag
\\
\phantom{\tilde{\sigma}(3;\underline{z},\underline{\dot{z}})}{}
=\sum\nolimits_{nk\ell}^{\prime}\bigg(\frac{\dot{z}_{n}}{\den(z_{n},z_{k},z_{\ell})}
 \notag
\\
\phantom{\tilde{\sigma}(3;\underline{z},\underline{\dot{z}})=}{}
\times
\big\{(z_{k}-z_{\ell})\big[z_{n}^{2}-(z_{n}-z_{k})(z_{n}-z_{\ell})-z_{k}^{2}\big]-z_{\ell}^{2}(z_{\ell}+z_{k}
-2z_{n})\big\}\bigg)\notag
\\
\phantom{\tilde{\sigma}(3;\underline{z},\underline{\dot{z}})}{}
=\sum\nolimits_{nk\ell}^{\prime}\bigg\{\dot{z}_{n}\frac{-(z_{k}^{3}+z_{\ell}^{3})+z_{n}(z_{k}^{2}+z_{\ell}
^{2})}{\den(z_{n},z_{k},z_{\ell})}\bigg\}=0.
\end{gather}
Here the f\/irst equality is justif\/ied by the def\/inition~\eqref{denden} and a~bit of trivial algebra,
the second equality obtains by trivial algebra, the third equality obtains by the exchange of the dummy
indices $n\leftrightarrow k$ in the term multiplying~$\dot{z}_{k}$ and likewise the exchange
$n\leftrightarrow \ell$ in the term multiplying $\dot{z}_{\ell}$ (under these exchanges the denominator
changes sign, see~\eqref{denanti}), the fourth equality obtains by trivial algebra, and the f\/inal
equality to zero is yielded by  Lemma~\ref{LemmaA.1} since the numerator is invariant under the exchange
$k\leftrightarrow \ell$.

\subsection*{Acknowledgements}

One of us (OB) would like to acknowledge with thanks the hospitality of the Physics Department of the
University of Rome ``La Sapienza'' on the occasion of two two-week visits in June 2012 and May 2013, and
the f\/inancial support for these trips provided by the NSF-AWM Travel Grant.

\pdfbookmark[1]{References}{ref}
\LastPageEnding


\begin{thebibliography}{99}
\footnotesize\itemsep=0pt

\bibitem{BCCircle13}
Bihun O., Calogero F., Solvable and/or integrable many-body models on a circle,
  \textit{J.~Geom. Symmetry Phys.} \textbf{30} (2013), 1--18.

\bibitem{BCY13}
Bihun O., Calogero F., Yi G., Diophantine properties associated to the
  equilibrium conf\/igurations of an isochronous $N$-body problem,
  \href{http://dx.doi.org/10.1080/14029251.2013.792494}{\textit{J.~Nonlinear Math. Phys.}} \textbf{20} (2013), 158--178.

\bibitem{C1978}
Calogero F., Motion of poles and zeros of special solutions of nonlinear and
  linear partial dif\/ferential equations and related ``solvable'' many-body
  problems, \href{http://dx.doi.org/10.1007/BF02721013}{\textit{Nuovo Cimento~B}} \textbf{43} (1978), 177--241.

\bibitem{C2001}
Calogero F., Classical many-body problems amenable to exact treatments,
  \textit{Lecture Notes in Physics. New Series~m: Monographs}, Vol.~66,
  Springer-Verlag, Berlin, 2001.

\bibitem{Gold}
Calogero F., The neatest many-body problem amenable to exact treatments (a
  ``goldf\/ish''?), \href{http://dx.doi.org/10.1016/S0167-2789(01)00160-9}{\textit{Phys.~D}} \textbf{152--153} (2001), 78--84.

\bibitem{C2008}
Calogero F., Isochronous systems, \href{http://dx.doi.org/10.1093/acprof:oso/9780199535286.001.0001}{Oxford University Press}, Oxford, 2008.

\bibitem{C11}
Calogero F., An integrable many-body problem, \href{http://dx.doi.org/10.1063/1.3638052}{\textit{J.~Math. Phys.}}
  \textbf{52} (2011), 102702, 5~pages.

\bibitem{C12_1}
Calogero F., Another new goldf\/ish model, \href{http://dx.doi.org/10.1007/s11232-012-0060-3}{\textit{Theoret. and Math. Phys.}}
  \textbf{171} (2012), 629--640.

\bibitem{C12_2}
Calogero F., New solvable many-body model of goldf\/ish type,
  \href{http://dx.doi.org/10.1142/S1402925112500064}{\textit{J.~Nonlinear Math. Phys.}} \textbf{19} (2012), 1250006, 19~pages.

\bibitem{C12_3}
Calogero F., Two quite similar matrix {ODE}s and the many-body problems related
  to them, \href{http://dx.doi.org/10.1142/S021988781260002X}{\textit{Int.~J. Geom. Methods Mod. Phys.}} \textbf{9} (2012),
  1260002, 6~pages.

\bibitem{C13_2}
Calogero F., A linear second-order {ODE} with only polynomial solutions,
  \href{http://dx.doi.org/10.1016/j.jde.2013.06.007}{\textit{J.~Differential Equations}} \textbf{255} (2013), 2130--2135.

\bibitem{C13_3}
Calogero F., On the zeros of polynomials satisfying certain linear second-order
  ODEs featuring many free parameters, \href{http://dx.doi.org/10.1080/14029251.2013.805565}{\textit{J.~Nonlinear Math. Phys.}}
  \textbf{20} (2013), 191--198.

\bibitem{C2013}
Calogero F., A {\it solvable} many-body problem, its {\it equilibria}, and a
  second-order ordinary dif\/ferential equation whose {\it general} solution is
  polynomial, \href{http://dx.doi.org/10.1063/1.4773571}{\textit{J.~Math. Phys.}} \textbf{54} (2013), 012703, 13~pages.

\bibitem{CY12_3}
Calogero F., Yi G., A new class of solvable many-body problems, \href{http://dx.doi.org/10.3842/SIGMA.2012.066}{\textit{SIGMA}}
  \textbf{8} (2012), 066, 29~pages, \href{http://arxiv.org/abs/1210.0651}{arXiv:1210.0651}.

\bibitem{CY12_1}
Calogero F., Yi G., Can the {\it general} solution of the second-order {ODE}
  characterizing {J}acobi polynomials be {\it polynomial}?, \href{http://dx.doi.org/10.1088/1751-8113/45/9/095206}{\textit{J.~Phys.~A:
  Math. Theor.}} \textbf{45} (2012), 095206, 4~pages.

\bibitem{CY12_2}
Calogero F., Yi G., Diophantine properties of the zeros of certain {L}aguerre
  and para-{J}acobi polynomials, \href{http://dx.doi.org/10.1088/1751-8113/45/9/095207}{\textit{J.~Phys.~A: Math. Theor.}} \textbf{45}
  (2012), 095207, 9~pages.


\bibitem{CY12_4}
Calogero F., Yi G., Polynomials satisfying functional and dif\/ferential
  equations and {D}iophantine properties of their zeros, \href{http://dx.doi.org/10.1007/s11005-013-0612-y}{\textit{Lett. Math.
  Phys.}} \textbf{103} (2013), 629--651.

\bibitem{GS2005}
Gomez-Ullate D., Sommacal M., Periods of the goldf\/ish many-body problem,
  \href{http://dx.doi.org/10.2991/jnmp.2005.12.s1.28}{\textit{J.~Nonlinear Math. Phys.}} \textbf{12} (2005), suppl.~1, 351--362.

\end{thebibliography}
\end{document}